Twelve- and fourteen-year-old school children differentially benefit from sensorimotor- and multisensory-enriched vocabulary training

*†Brian Mathias[1,2], †Christian Andrä[3,4], Anika Schwager[5],
Manuela Macedonia[2,6,7], & Katharina von Kriegstein[1,2]

†Joint first authors

[1]Chair of Cognitive and Clinical Neuroscience, Faculty of Psychology, Technical University Dresden, Dresden, Germany
[2]Max Planck Research Group Neural Mechanisms of Human Communication, Max Planck Institute for Human Cognitive and Brain Sciences, Leipzig, Germany
[3]Department of Teacher Education and School Research, University of Leipzig, Leipzig, Germany
[4]Department of School Sport, Institute of Sports Psychology and Physical Education, Faculty of Sports Science, University of Leipzig, Leipzig, Germany
[5]Didactics of Physical Education, Institute of Primary and Pre-Primary Education, Faculty of Education, University of Leipzig, Leipzig, Germany
[6]Institute for Information Engineering, Johannes Kepler University Linz, Linz, Austria
[7]Centre for Business Languages and Intercultural Communication, Johannes Kepler University Linz, Linz, Austria

*Corresponding author:
Brian Mathias
Technical University Dresden
Chair of Cognitive and Clinical Neuroscience
Faculty of Psychology
Bamberger Str. 7
01187 Dresden
Germany
Email: brian.mathias@tu-dresden.de

**Abstract**

Both children and adults have been shown to benefit from the integration of multisensory and sensorimotor enrichment into pedagogy. For example, integrating pictures or gestures into foreign language (L2) vocabulary learning can improve learning outcomes relative to unisensory learning. However, whereas adults seem to benefit to a greater extent from sensorimotor enrichment such as the performance of gestures in contrast to multisensory enrichment with pictures, this is not the case in elementary school children. Here, we compared multisensory- and sensorimotor-enriched learning in an intermediate age group that falls between the age groups tested in previous studies (elementary school children and young adults), in an attempt to determine the developmental time point at which children's responses to enrichment mature from a child-like pattern into an adult-like pattern. Twelve-year-old and fourteen-year-old German children were trained over 5 consecutive days on auditorily-presented, concrete and abstract, Spanish vocabulary. The vocabulary was learned under picture-enriched, gesture-enriched, and non-enriched (auditory-only) conditions. The children performed vocabulary recall and translation tests at 3 days, 2 months, and 6 months post-learning. Both picture and gesture enrichment interventions were found to benefit children's L2 learning relative to non-enriched learning up to 6 months post-training. Interestingly, gesture-enriched learning was even more beneficial than picture-enriched learning for the fourteen-year-olds, while the twelve-year-olds benefitted equivalently from learning enriched with pictures and gestures. These findings provide evidence for opting to integrate gestures rather than pictures into L2 pedagogy starting at fourteen years of age.

*Keywords:* multisensory learning, sensorimotor learning, gesture, enrichment, vocabulary learning, foreign language education

## Introduction

Modern classrooms often make use of multisensory learning materials (Choo, Lin, & Pandian, 2012; Kiefer & Trumpp, 2012). One reason for doing so is that presence of complementary information across multiple sensory and motor modalities may speed up learning and make it more resistant to decay (Mahmoudi, Jafari, Nasrabadi, & Liaghatdar, 2012; Sadoski & Paivio, 2013; Shams & Seitz, 2008; von Kriegstein & Giraud, 2006). For example, children tend to benefit more from visual grapheme training when it is integrated with auditory phonological training (reviewed in Ehri et al., 2001). Writing letters by hand can also benefit children's learning above and beyond unisensory visual training (Zemlock, Vinci-Booher, & James, 2018). Congruent information presented across two or more sensory modalities during learning has been referred to as *multisensory enrichment* (Mayer, Yildiz, Macedonia, & von Kriegstein, 2015), and the combination of body movements with information presented in one or more sensory modalities during learning has been referred to *sensorimotor enrichment* (reviewed in Macedonia, 2014).

Foreign language (L2) learning is one domain that stands to benefit from enriched classroom instruction. One of the most prevalent means of learning L2 vocabulary is students' use of written word lists (Oxford & Crookall, 1990; Schmitt & Schmitt, 2020). However, recent work has suggested that multisensory enrichment can boost L2 vocabulary acquisition. Silverman and Hines (2009), for example, found that the viewing of short video clips that supplemented teachers' regular instruction improved kindergartners' through second graders' acquisition of L2 vocabulary. The video clips were excerpts of documentaries, such as National Geographic's *Really Wild Animals* series (National Geographic, 2005), that contained target L2 words. Other studies have suggested benefits of flash cards (Li & Tong, 2019) and pictures paired with audio recordings (Andrä, Mathias, Schwager, Macedonia, & von Kriegstein, 2020). Further work has provided evidence for benefits of sensorimotor enrichment on L2 vocabulary learning. In one study, students' performance of iconic gestures in tandem with physical exercise while listening to foreign language vocabulary increased recall compared to exercising without gestures (Mavilidi, Okely, Chandler, Cliff, & Paas, 2015). Spanish ten-year-olds' comprehension of stories told in English improved if the instructor enacted gestures during the story-telling (Cabrera & Martinez, 2001), and German high school students' memory for Latin words benefitted from the integration of choral speech and meaningful gestures and movements into the memorization process (Hille, Vogt, Fritz, & Sambanis, 2010). Finally, Macedonia, Bergmann, and Roithmayr (2014; see also de Wit et al., 2018) demonstrated that eleven-year-old children's L2 vocabulary learning outcomes were aided more by performing semantically-related gestures themselves during learning than by viewing a pedagogical agent perform the gestures.

    Benefits of multisensory and sensorimotor enrichment in these studies can be explained in terms of embodied memory for L2 words (reviewed in Atkinson, 2010), dual coding of L2 word representations (Engelkamp & Zimmer, 1984; Hommel, Müsseler, Aschersleben, & Prinz, 2001; Paivio, 1991; Paivio & Csapo, 1969), mental imagery of multimodally-represented L2 words (Jeannerod, 1995; Kosslyn, Thompson, & Ganis, 2006; Saltz & Dixon, 1982), and predictive coding accounts of L2 representations (Mayer et al., 2017; Mathias et al., 2020; von Kriegstein 2012). A common thread of these accounts is that novel information can be mentally represented in terms of its perceptual and motor features, which may aid learning and memory. At a neural level, the same sensory and motor brain regions that process visuomotor enrichment information during learning appear to be causally relevant for subsequent auditory L2 recognition (Mayer et al., 2015; Mathias et al., 2020). The notion that brain regions that support the processing of enrichment also drive enrichment-based learning benefits has been referred to as multisensory learning theory (von Kriegstein, 2012).

    A key question for the development of evidence-based teaching strategies is whether multisensory enrichment techniques are more (or less) effective than sensorimotor enrichment techniques. This question is of interest in light of growing support for the effectiveness of active learning techniques in educational settings, defined as instructional methods that engage students in the learning process (Drew & Mackie, 2011; Jensen, Kummer, & Godoy, 2015;

Michael, 2006; Prince, 2004; Sambanis, 2013). One recent study directly compared effects of a multisensory enrichment technique (learning with pictures) with sensorimotor-enriched learning (learning with gestures) in the context of L2 vocabulary learning (Andrä et al., 2020). In this study, both picture-enriched and gesture-enriched learning enhanced eight-year-olds' free recall and translation of L2 words compared to a unisensory learning baseline condition. Benefits of picture and gesture enrichment were approximately equivalent, even up to 6 months after the L2 vocabulary instruction had ended. This finding in children contrasts with findings in adults in laboratory environments. Adults' L2 vocabulary learning has been shown to benefit more from performing gestures during learning than viewing pictures (Mathias et al., 2020; Mayer et al., 2015). This effect is particularly pronounced over the long-term (several months post-learning), suggesting that picture-enriched L2 words decay more quickly from memory than gesture-enriched L2 words.

      The discrepancy between findings in children and adults with regard to enriched learning strategies suggests that teaching strategies derived from studies on adults or vice versa may not directly translate into teaching strategies for children. Some studies have revealed learning mechanisms that are highly similar across children and adults, such as auditory statistical learning, which remains relatively constant through the course of development (Raviv & Arnon, 2016; Saffran, Johnson, Aslin, & Newport, 1999). However, children and adults are also known to differ with regard to several key learning mechanisms, such as their use of working memory (Luna, Garver, Urban, Lazar, & Sweeney, 2004) and deployment of visual and motor imagery (Frick et al., 2009; Funk, Brugger, & Wilkening, 2005; reviewed in Gabbard, 2009). Differences in enrichment effects of different age groups also have immediate implications for evidence-based teaching techniques, as gestures and other sensorimotor-based interventions may be more challenging for educators to integrate into pedagogy than multisensory-based interventions.

      In the present study, we compare multisensory- and sensorimotor-enriched learning in an age group that falls between the age groups tested in previous studies (elementary school

children and young adults). The sample included both twelve-year-olds (sixth graders) and fourteen-year-olds (eighth graders) who were all currently enrolled in their first semester of learning Spanish as a foreign language. Our aim was to test whether differences in effects of multisensory (picture) and sensorimotor (gesture) enrichment previously observed in adults (Mathias et al., 2020; Mayer et al., 2015), but not in elementary school children, occur for this intermediate-aged group of high school children. We hypothesized that, if high school children are more similar to elementary school children in terms of their response to picture and gesture enrichment (Andrä et al., 2020), then we would observe no differences between effects of the two enrichment types. However, if high school children are more similar to young adults, then we would observe a greater benefit of gesture enrichment compared to picture enrichment. A third possibility was that the pattern of enrichment effects might diverge across age groups, i.e., twelve-year-olds would show equivalent picture and gesture benefits, and fourteen-year-olds would show a greater gesture than picture benefit.

Besides testing our main hypotheses outlined above, we expected three further effects that have already been shown in adults (Macedonia & Knösche, 2011; Mayer et al., 2015; Mayer et al., 2017; Repetto, Pedroli, & Macedonia, 2017) and elementary school children (Andrä et al., 2020). First, we expected that high-school-aged children would demonstrate benefits of picture- and gesture-enriched learning compared to non-enriched (auditory-only) learning. Second, we expected the beneficial effects of picture and gesture enrichment to persist over long time scales (up to 6 months following learning; Andrä et al., 2020; Mayer et al., 2015). We therefore tested the high school children's knowledge of the enriched vocabulary at three different time points: 3 days, 2 months, and 6 months post-learning. Finally, we expected that both picture and gesture enrichment would benefit high school children's learning of both concrete (e.g., *tent*) and abstract words (e.g., *patience*) compared to non-enriched learning.

**Methods**

**Participants**

Participants were school children enrolled in Spanish foreign language courses at three public high schools located in the vicinity of Chemnitz, Germany. Forty-eight children were enrolled in grade 6 (twelve- to thirteen-year-olds) and 47 children were enrolled in grade 8 (fourteen- to fifteen-year-olds). Regardless of their grade level (grade 6 or grade 8), all children were currently enrolled in their first course of Spanish as a foreign language and had not previously received any Spanish language training or lessons. Written informed consent was obtained from the legal guardians of all school children who participated. The investigators briefed the children and their teachers on the study procedures in an introductory session that took place prior to the experiment. Children who were absent from at least one training or test session were excluded from the analyses. Therefore, the analyses included 39 children in grade 6 ($M$ age = 12.8 years, $SD$ = 0.4 years, 20 females) and 36 children in grade 8 ($M$ age = 14.8 years, $SD$ = 0.4 years, 27 females). Based on the teachers' reports, none of the children possessed learning disabilities, and all of the children possessed normal or corrected-to-normal vision. Two of the children in grade 8 and none of the children in grade 6 spoke another language besides German and English. The study was reviewed and approved by the Education Department of the state of Saxony, Germany.

**Stimulus Materials**

Spanish words used in the experiment were selected in consultation with the children's school teachers at each of the three high schools. Word selection was based on three factors: First, children had not yet encountered the words in lessons and the words were not anticipated to be included in the teaching curriculum for the 6-month duration of the investigation. Second, the words were considered by the teacher to be relevant for future use by the children. Third, words were among the 90 words included in the "Vimmi" language corpus (Macedonia, Müller, & Friederici, 2010, 2011). The Vimmi corpus was created for experiments on L2 learning and contains videos of gestures designed to convey the meanings of words included in the corpus. This resulted in one set of 24 Spanish words for each of the three high schools, shown in **Table 1**.

|  | **High School 1** |  |  |  |  |
|---|---|---|---|---|---|
| **Concrete nouns** |  |  | **Abstract nouns** |  |  |
| German | Spanish | English | German | Spanish | English |
| Anhänger | colgante | trailer | Absage | negativa | cancellation |
| Bildschirm | pantalla | monitor | Anforderung | exigencia | requirement |
| Briefkasten | buzón | letter box | Aufwand | esfuerzo | effort |
| Faden | hilo | thread | Besitz | propiedad | property |
| Kabel | cable | cable | Bestimmung | destinación | destination |
| Katalog | catálogo | catalog | Geduld | paciencia | patience |
| Maske | máscara | mask | Gleichgültigkeit | indiferencia | indifference |
| Reifen | neumático | tire | Rücksicht | respeto | consideration |
| Ring | anillo | ring | Tatsache | hecho | fact |
| Sammlung | colección | collection | Vorwand | pretexto | excuse |
| Verband | vendaje | bandage | Wohltat | beneficio | benefaction |
| Zelt | tienda | tent | Zulassung | admisión | admission |
|  | **High School 2** |  |  |  |  |
| **Concrete nouns** |  |  | **Abstract nouns** |  |  |
| German | Spanish | English | German | Spanish | English |
| Ampel | semáforo | traffic light | Absage | negativa | cancellation |
| Anhänger | colgante | trailer | Anforderung | exigencia | requirement |
| Balkon | balcón | balcony | Aufwand | esfuerzo | effort |
| Bildschirm | pantalla | monitor | Bestimmung | destinación | destination |
| Faden | hilo | thread | Geduld | paciencia | patience |
| Kabel | cable | cable | Gleichgültigkeit | indiferencia | indifference |
| Katalog | catálogo | catalog | Mut | valor | bravery |
| Maske | máscara | mask | Rücksicht | respeto | consideration |
| Reifen | neumático | tire | Verständnis | sensibilidad | sympathy |
| Schublade | cajón | ring | Vorwand | pretexto | excuse |
| Verband | vendaje | bandage | Wohltat | beneficio | benefaction |
| Zigarette | cigarrillo | cigarette | Zulassung | admisión | admission |
|  | **High School 3** |  |  |  |  |

|  | **Concrete nouns** |  |  | **Abstract nouns** |  |
|---|---|---|---|---|---|
| German | Spanish | English | German | Spanish | English |
| Ampel | semáforo | traffic light | Befehl | orden | command |
| Briefkasten | buzón | letter box | Bitte | ruego | plea |
| Denkmal | monumento | memorial | Empfehlung | recomendación | recommendation |
| Fernbedienung | telemando | remote control | Gedanke | pensamiento | thought |
| Flugzeug | avión | airplane | Korrektur | corrección | correction |
| Handtasche | bolso de mano | purse | Mentalität | mentalidad | mentality |
| Kabel | cable | cable | Mut | valor | bravery |
| Koffer | maleta | suitcase | Talent | talento | talent |
| Schlüssel | llave | key | Teilnahme | participación | participation |
| Straßenbahn | tranvía | tram | Tradition | tradición | tradition |
| Tageszeitung | diario | daily newspaper | Veränderung | cambio | change |
| Zelt | tienda | tent | Warnung | advertencia | warning |

**Table 1.** Vocabulary used at each of the three high schools included in the experiment. Twenty-four German and Spanish words were used at each high school. English translations are also shown. Assignment of words to the gesture enrichment, picture enrichment, and no enrichment conditions was counterbalanced across participants at each school, ensuring that each German and Spanish word was represented equally in each of the learning conditions.

Half of the words used at each school were concrete nouns and the other half were abstract nouns. Concreteness and imageability ratings (on a 0 to 10 scale) derived from a corpus of German lemmas (Köper & Schulte im Walde, 2016) are shown in **Table 2**. Imageability refers to the ease with which a word gives rise to a sensorimotor mental image (Paivio, 1971). Concreteness and imageability ratings were significantly higher for the concrete words compared to abstract words for each of the three schools (all $p$s < .001; **Table 3**). Abstract and concrete word frequencies in written German (http://wortschatz.uni-leipzig.de/en) did not significantly differ, shown in **Table 3**.

|  | **Concrete nouns** | **Abstract nouns** |
|---|---|---|

| Word | Concrete-ness | Image-ability | Word | Concrete-ness | Image-ability |
|---|---|---|---|---|---|
| airplane | 7.7 | 7.6 | admission | 3.2 | 4.2 |
| balcony | 7.0 | 7.0 | benefaction | 2.1 | 3.5 |
| bandage | 5.2 | 5.2 | bravery | 3.0 | 4.4 |
| cable | 6.5 | 6.6 | cancellation | 3.4 | 3.6 |
| catalog | 5.0 | 6.0 | change | 2.9 | 2.9 |
| cigarette | 7.9 | 7.4 | command | 4.0 | 3.8 |
| collection | 4.1 | 4.5 | consideration | 2.4 | 2.8 |
| daily | 6.3 | 6.9 | correction | 4.4 | 3.4 |
| key | 6.2 | 6.0 | destination | 2.5 | 2.9 |
| letter box | 6.9 | 6.1 | effort | 2.2 | 2.5 |
| mask | 6.4 | 6.6 | excuse | 3.4 | 3.8 |
| memorial | 5.8 | 5.9 | fact | 2.0 | 1.9 |
| monitor | 6.7 | 6.5 | indifference | 2.1 | 3.5 |
| purse | 7.7 | 7.1 | mentality | 1.9 | 2.6 |
| remote control | 6.2 | 5.7 | participation | 3.7 | 3.8 |
| ring | 7.1 | 7.1 | patience | 2.1 | 3.5 |
| ring | 6.1 | 5.5 | plea | 4.6 | 4.3 |
| suitcase | 7.5 | 7.9 | property | 3.1 | 4.1 |
| tent | 7.6 | 7.9 | recommendation | 3.3 | 3.0 |
| thread | 6.2 | 6.0 | requirement | 1.8 | 2.3 |
| tire | 6.8 | 6.0 | sympathy | 1.8 | 2.5 |
| traffic light | 6.7 | 6.5 | talent | 3.4 | 3.7 |
| trailer | 5.7 | 5.5 | thought | 2.9 | 3.8 |
| tram | 6.6 | 7.2 | tradition | 2.3 | 2.7 |
|  |  |  | warning | 3.4 | 3.7 |

**Table 2.** Concreteness and imageability ratings of the words used in the experiment (derived from Köper & Schulte im Walde, 2016).

|  | **High School 1** | | | | | | |
|---|---|---|---|---|---|---|---|
|  | **Concrete nouns** | | **Abstract nouns** | | | | |
|  | *M* | *SD* | *M* | *SD* | *t* | *p* | *d* |
| Concreteness | 6.2 | 1.0 | 2.5 | .6 | 10.90 | <.001*** | 4.5 |
| Imageability | 6.2 | .9 | 3.2 | .7 | 8.92 | <.001*** | 3.7 |
| Frequency | 11.0 | 1.2 | 11.2 | 1.2 | .34 | .74 | .2 |
|  | **High School 2** | | | | | | |
|  | **Concrete nouns** | | **Abstract nouns** | | | | |
|  | *M* | *SD* | *M* | *SD* | *t* | *p* | *d* |

|               | M    | SD  | M    | SD  | t     | p        | d   |
|---------------|------|-----|------|-----|-------|----------|-----|
| Concreteness  | 6.4  | .8  | 2.5  | .6  | 13.42 | <.001*** | 5.6 |
| Imageability  | 6.2  | .7  | 3.3  | .7  | 10.78 | <.001*** | 4.3 |
| Frequency     | 11.1 | 1.1 | 11.2 | 1.2 | .18   | .86      | .1  |

|               | **High School 3** | | | | | | |
|---------------|------|-----|------|-----|-------|----------|-----|
|               | **Concrete nouns** | | **Abstract nouns** | | | | |
|               | M    | SD  | M    | SD  | t     | p        | d   |
| Concreteness  | 6.8  | .7  | 3.3  | .8  | 11.68 | <.001*** | 5.1 |
| Imageability  | 6.8  | .8  | 3.5  | .6  | 11.63 | <.001*** | 4.9 |
| Frequency     | 11.7 | 1.0 | 11.2 | .9  | 1.27  | .22      | .5  |

**Table 3.** Concreteness ratings, imageability ratings, and frequencies for the concrete and abstract German words used in the experiment at each of the three high schools. *df* = 22 for all *t* tests. ***p* < .001.

The experiment made use of three stimulus types: audio recordings of Spanish words and their German translations, pictures depicting word meanings, and videos of an actress performing gestures that were semantically related to word meanings. Audio recordings of German words, as well as picture and video stimuli, were adopted from the Vimmi corpus (Macedonia et al., 2010, 2011; Mayer et al., 2015).

The German word recordings featured a female bilingual Italian-German speaker (age 44). Recordings of Spanish translations featured a female native speaker of European Spanish (age 25). Recordings were made using a RØDE NT55 microphone (RØDE Microphones, Silverwater, Australia) in a sound-dampened room.

The pictures consisted of black-and-white line drawings created by a professional cartoon artist (https://www.klaus-pitter.com/). The drawings iconically communicated word meanings by depicting objects, humans, or scenes. Abstract nouns were conveyed using scenes. Pictures representing one of the concrete nouns and one of the abstract nouns are shown in **Figure 1**.

The complexity of line drawings was not matched for concrete and abstract nouns, as differences in complexity are also expected to occur in naturalistic teaching settings.

Videos were recorded using a Canon Legria HF S10 camcorder (Canon Inc., Tokyo, Japan). Each video was 4 s long and shot in color. The actress shown in the videos began and ended each video by standing motionless with her arms by her sides. During the videos, she used head movements, movements of one or both arms or legs, fingers, or combinations of these body parts to convey the meaning of the foreign language word through the movement. For example, the word *tent* was conveyed by moving the arms and fingers together to form an upside-down "V" shape, and the word *patience* was conveyed by lifting up the arms and subsequently slowly moving them outward from the body and downward (**Figure 1**). The actress always maintained a neutral facial expression. Gestures selected for abstract nouns were previously agreed upon by three independent raters (Macedonia et al., 2011; Mayer et al., 2015).

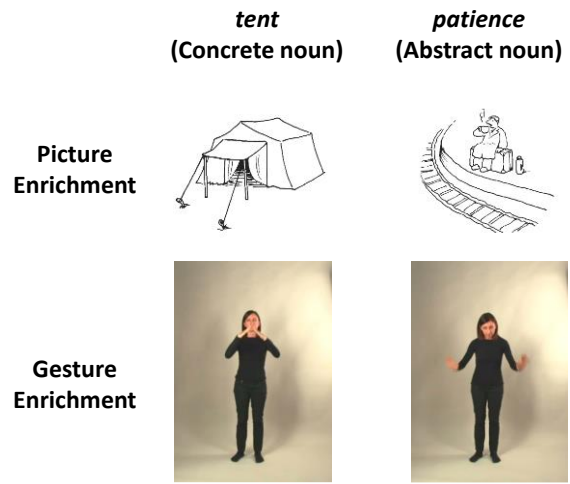

**Figure 1.** Picture and gesture stimuli. Top: Pictures used in the picture enrichment condition for one of the concrete nouns (*tent*) and one of the abstract nouns (*patience*). Bottom: Screen captures from the corresponding videos of the actress performing gestures, which were used in the gesture enrichment condition.

**Design and Procedure**

The experiment had a 3 × 2 × 3 × 2 factorial within-participants design with the factors learning condition (gesture enrichment, picture enrichment, no enrichment), grade level (grade 6, grade 8), testing time point (3 days, 2 months, and 6 months post-learning), and word type (concrete, abstract).

**Learning phase.** Children completed L2 vocabulary training that took place over a period of 8 days (**Figure 2a**). Training was integrated within children's regular Spanish course meetings, and therefore took place on day 1 for 90 mins, day 3 or 4 for 45 mins, and day 8 for 90 mins. The second training session occurred on either day 3 or day 4 because of differences in Spanish course scheduling between schools.

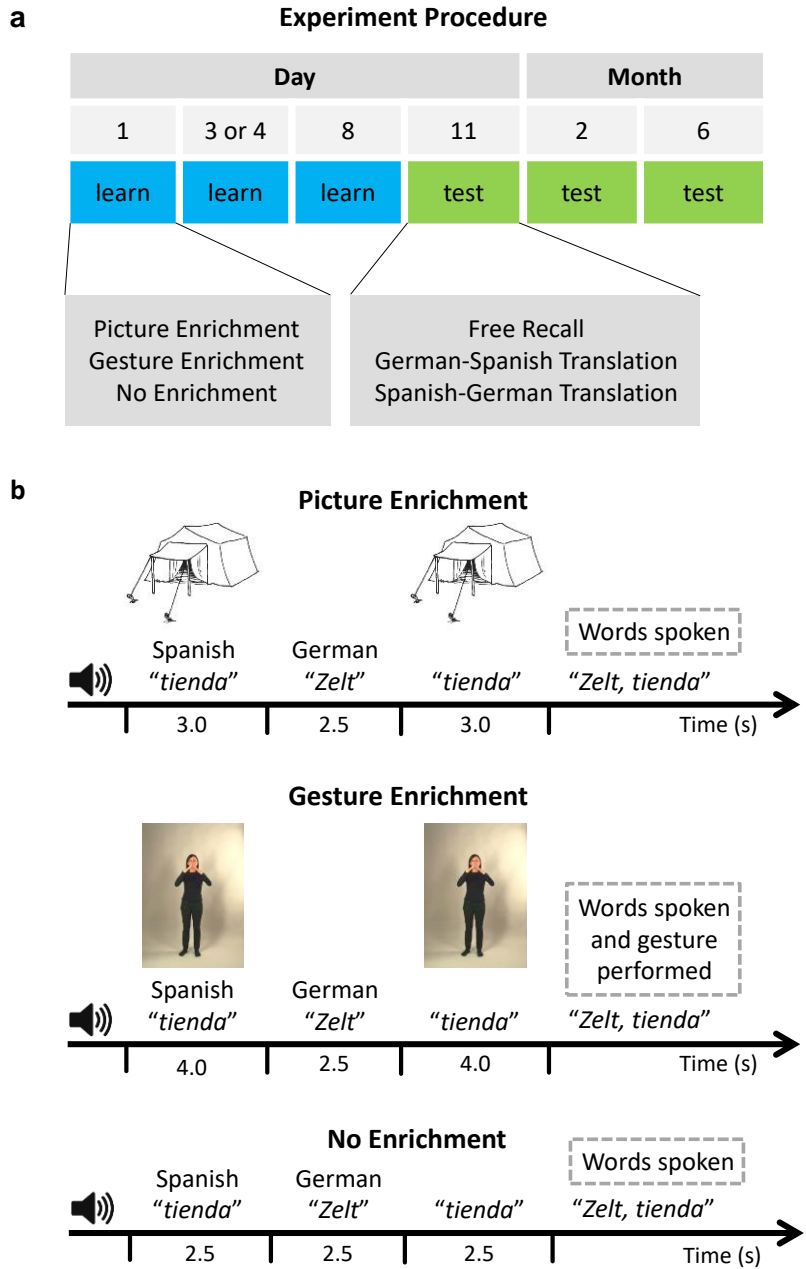

**Figure 2.** Experimental procedure and design. **(a)** The learning phase of each experiment occurred over 8 days ('learn'). Free recall and translation tests ('test') were administered 3 days, 2 months, and 6 months following the end of the learning phase. High school children learned foreign language words in gesture, picture, and no enrichment conditions. **(b)** In each learning trial, auditorily-presented Spanish words were accompanied either by a picture (picture enrichment), a video of an actress performing a gesture (gesture enrichment), or no complimentary stimulus (no enrichment). Spanish words were followed by the auditorily-presented German translation and a repetition of the Spanish word accompanied again by the enrichment stimulus. The children then spoke the foreign and native words following their teacher. In the gesture enrichment condition, the children performed gestures with their teacher while speaking. The children's task was to learn the correct association between the Spanish words and their German translations.

During the training sessions, L2 words and their L1 translations were presented in picture-enriched, gesture-enriched trials, and non-enriched trials (**Figure 2b**). In all trial types, children first heard a Spanish word, which was followed by its auditorily-presented German translation and then by a repetition of the Spanish word. The children's teacher then cued the children to recite the German and Spanish words aloud with the word *juntos*, which means *all together*. The teacher stood at the front of the classroom during the entire training period. In the picture enrichment condition, recorded Spanish words were accompanied by iconic line drawings, which were presented on a screen at the front of the classroom. Pictures were presented for 3 s. In the gesture enrichment condition, recorded Spanish words were accompanied by videos of an actress performing an iconic gesture, which lasted 4 s. At the end of the trial, children performed the gesture along with the teacher. The time interval between the onset of the German word's presentation and the onset of the Spanish word's repetition was 2.5 s in all three learning conditions. In order to equate the time interval between the offset of the pictures or videos and the subsequent German word onset, and to allow for comparison with previous experiments (Andrä et al., 2020), the time interval between Spanish and German word onsets in the non-enriched learning condition was set to 2.5 s. Children's locations in the classroom were randomly assigned for each training block. Children sat at desks during the non-enriched and picture-enriched trials and stood next to desks during the gesture-enriched trials. One of the investigators monitored the testing equipment and initiated each trial as soon as the children were ready.

Learning phase trials were blocked by learning condition. Each block contained 8 trials (4 concrete word trials and 4 abstract word trials) and lasted approximately 4 mins. On day 1, the children completed 2 picture-enriched blocks, 2 gesture-enriched blocks, and 2 non-enriched blocks. Each German word and its Spanish translation were therefore presented in two trials on day 1. On day 3 or day 4, the children completed 1 picture-enriched block, 1 gesture-enriched block, and 1 non-enriched block. Fewer blocks were administered on day 3 or 4 compared to other days due to the shorter Spanish course meeting time on day 3 or 4. On day 8, the children completed 2 picture-enriched blocks, 2 gesture-enriched blocks, and 2 non-enriched blocks. Children rested in a separate room between every two blocks for approximately 5 minutes, during which time they played simple riddle games with one of the experimenters.

Children were equally divided into groups of up to 9 students in order to counterbalance the assignment of word stimuli to the three learning conditions. This ensured that each stimulus item was learned by students in each of the three learning conditions, and that stimuli did not vary systematically between learning conditions. Additionally, word orders within each block and orders of enrichment condition blocks were counterbalanced across learning days.

**Test phase.** Children completed vocabulary tests at three time points: 3 days, 2 months, and 6 months following the completion of the learning phase. Free recall, German-Spanish, and Spanish-German translation tests were conducted at each time point. Tests were conducted entirely verbally, since the children did not yet possess adequate writing skills in Spanish as a foreign language.

Native German-speaking examiners conducted the test sessions individually at the same school where the learning phase took place. The examiners were university students enrolled in teaching certification programs at the University of Leipzig, Germany. Examiners were blind with respect to which words had been learned in which enrichment condition. Further, they had no knowledge of the gestures or pictures that were paired with individual words in the experiment.

During each test session, one of the school children sat at a desk opposite one of the examiners. In the free recall test, children were asked to verbalize as many German-Spanish or Spanish-German translations, individual German words, or individual Spanish words as they could remember from the training. A time limit of 5 minutes was imposed; children were not instructed about this time limit, and no child's responses in any experiment exceeded 5 mins. Following the free recall test, the children completed the two translation tests. The free recall test was always administered prior to the translation tests to eliminate influences of memory cues present in the translation tests.

During the German-Spanish translation test, the examiner spoke the German words one at a time, and the children were asked each time to speak the correct Spanish translation. During the Spanish-German translation test, the examiner presented audio recordings of the Spanish words one at a time, and the children were asked each time to speak the correct German translation. The German-Spanish translation test was always administered prior to the Spanish-German test, as translation from one's native to a foreign language has been shown to be a more difficult task than the translation from a foreign language into one's native language (Kroll & Stewart, 1994). Children were given 5 s to state their answers before moving to the next word. Test word orders in the two translation tests were randomized for each testing time point (3 days, 2 months, and 6 months post-learning).

Examiners recorded test sessions as an audio file for subsequent analysis using a recording device such as mobile phones or laptops. The children did not receive any feedback regarding the correctness of their answers. Children were instructed not to discuss the tests with their classmates. Each test session lasted approximately 10-15 minutes.

**Data Analysis**

Audio files from individual test sessions were independently scored for accuracy by two raters. The raters were a Spanish-German bilingual speaker and a native German speaker who were both currently enrolled in the Spanish language teaching certification program at the University of Leipzig. The two raters had not conducted any of the test sessions and were also blind with respect to which words had been learned in each enrichment condition. In cases of disagreement, a third independent rater was employed and the majority decision was adopted. The third rater was also a native German speaker currently enrolled in the Spanish language teaching certification program at the University of Leipzig.

One point was given for each correct translation provided during the free recall test. No points were given for a German word that was missing a corresponding Spanish translation or vice versa. One point was also given for each correct translation provided in the German-Spanish translation test and the Spanish-German translation test. Scores across the three tests were summed for each participant, yielding combined test scores for each experimental condition.

Linear mixed effects models were used to evaluate effects of enrichment condition, grade level, time point, and vocabulary type on learning outcomes. Mixed effects models were generated in R version 1.2.1335 using the „lme4" package (Bates, Maechler, Bolker, & Walker, 2015), and were generated separately for data from each vocabulary test (L1-L2 translation, L2-L1 translation, and free recall). All mixed effects models included fixed effects of enrichment type (gesture, picture, none), grade level (6, 8), time point (3 days post-learning, 2 months post-learning, 6 months post-learning), and word type (concrete, abstract). To select the random effects structure, we performed backwards model selection, beginning with a random intercept by participant and random slopes by participant for each of the four independent factors (enrichment type, grade level, time point, and word type). We removed random effects terms that accounted for the least variance one by one until the fitted mixed model was no longer singular, i.e., until variances of one or more linear combinations of random effects were no longer (close to) zero. The final mixed models for each vocabulary test included two random effects terms: a random intercept by participant and a random slope by participant for the word type factor.

Model contrasts were coded using simple coding, i.e., ANOVA-style coding, such that the model coefficient represented the size of the contrast from a given predictor level to the (grand) mean (represented by the intercept). The complete set of mixed effects model coefficient estimates for all fixed and random effects is shown in supplementary **Table S1**. Following the procedure outlined by Alday et al. (2017), significance testing of effects was performed using Type-II Wald $\chi^2$ tests implemented in the 'car' package (function: Anova(); Fox & Weisberg, 2011). Post-hoc Tukey tests were conducted using the 'emmeans' package (Lenth, Singmann, Love, Buerkner, & Herve, 2019). The significance threshold was set to $\alpha = 0.05$ (Greenland et al., 2016).

## Results

We first addressed our main aim of the paper by testing whether twelve-year-olds (sixth graders) and fourteen-year-olds (eighth graders) benefited similarly or differently from gesture and picture enrichment. Mixed effects modeling of children's vocabulary test scores revealed an interaction between the children's grade level and the learning condition ($\chi^2$ (2, $N = 75$) = 5.70, $p = .04$; the complete set of model coefficients for all fixed and random effects is shown in supplementary **Table S1**, and significance testing of model effects is shown in **Table 4**). Tukey's HSD post-hoc tests showed that children in both grade levels benefitted from gesture enrichment relative to non-enriched learning (grade 6: $\beta = 1.56$, $t = 6.05$, $p < .001$, grade 8: $\beta = 1.87$, $t = 6.99$, $p < .001$), shown in **Figure 3**. This was also the case for the picture enrichment condition (grade 6: $\beta = 1.47$, $t = 5.70$, $p < .001$, grade 8: $\beta = .92$, $t = 3.42$, $p = .008$). However, gesture enrichment enhanced learning outcomes even more than picture enrichment for the eighth graders ($\beta = .95$, $t = 3.56$, $p = .005$), which was not the case for the sixth graders ($\beta = .09$, $t = .35$, $p = .99$). In sum, both groups of children benefitted from both types of enrichment, and gesture enrichment was even more beneficial than picture enrichment for the older children than the younger children.

|                                          | $\chi^2$ | df | p        |
|------------------------------------------|----------|----|----------|
| Learning                                 | 91.82    | 2  | <.001*** |
| Grade                                    | 2.24     | 1  | .13      |
| Vocabulary                               | 44.45    | 1  | <.001*** |
| Time                                     | 25.57    | 2  | <.001*** |
| Learning × Grade                         | 5.70     | 2  | .04*     |
| Learning × Vocabulary                    | 1.08     | 2  | .58      |
| Grade × Vocabulary                       | 90.30    | 1  | <.001*** |
| Learning × Time                          | 4.29     | 4  | .37      |
| Grade × Time                             | 13.15    | 2  | .001**   |
| Vocabulary × Time                        | .53      | 2  | .77      |
| Learning × Grade × Vocabulary            | .55      | 2  | .76      |
| Learning × Grade × Time                  | 3.41     | 4  | .49      |
| Learning × Vocabulary × Time             | 3.02     | 4  | .55      |
| Grade × Vocabulary × Time                | .24      | 2  | .89      |
| Learning × Grade × Vocabulary × Time     | 1.25     | 4  | .87      |

**Table 4.** Type-II Wald $\chi^2$ test of mixed effects model effects of learning condition, grade level, vocabulary type, and time point. *df* = degrees of freedom. *p* < .05, **p* < .01, ***p* < .001.

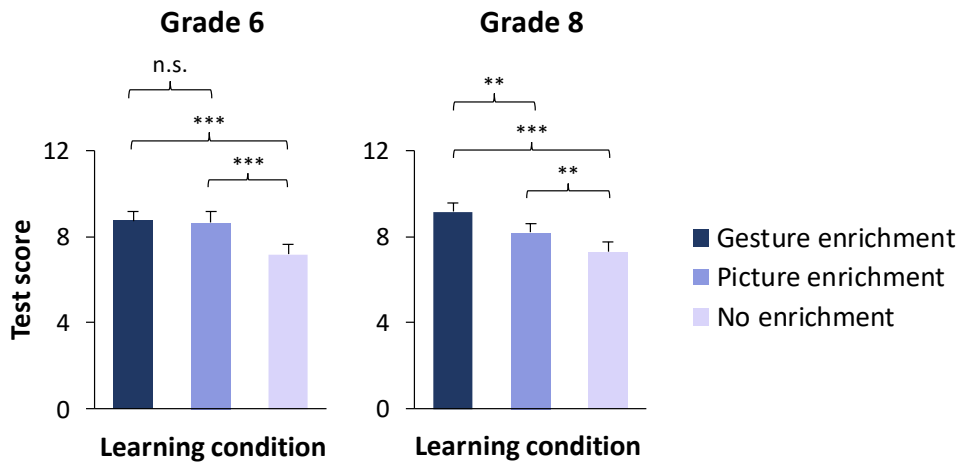

**Figure 3.** Test scores by learning condition and children's grade level. Children in grades 6 (twelve-year-olds; left) and grade 8 (fourteen-year-olds; right) demonstrated higher overall test scores following gesture-enriched learning and picture-enriched learning compared to non-enriched learning. Eighth graders benefitted significantly more from gesture enrichment than picture

enrichment, while sixth graders demonstrated equivalent learning outcomes for both gesture- and picture-enriched words. This difference was significant, i.e., there was an interaction between the learning condition and grade level factors. **$p$ < 0.01, ***$p$ < .001. n.s. = not significant.

We next tested whether gesture enrichment and picture enrichment would benefit children's test scores compared to non-enriched learning, irrespective of grade level, as expected from previous studies in elementary school children and adults (Andrä et al., 2020; Mayer et al., 2015). The mixed effects model indicated significantly higher scores for words learned with gesture and picture enrichment compared to words learned with no enrichment (gesture condition: $β$ = 1.71, $t$ = -9.23, $p$ < .001; picture condition: $β$ = 1.19, $t$ = 6.42, $p$ < .001). The model also revealed that, overall, scores for gesture-enriched words were significantly higher than scores for picture-enriched words ($β$ = .52, $t$ = 2.81, $p$ = .014).

We also expected the beneficial effects of picture and gesture enrichment on children's learning to persist over long time scales (up to 6 months following learning; Andrä et al., 2020; Mayer et al., 2015). This was found to be the case; there was no significant interaction between learning condition and time point ($χ^2$ (4, $N$ = 75) = 4.29, $p$ = .37). Both gesture- and picture-enriched learning benefitted children's L2 vocabulary learning outcomes compared with non-enriched learning, irrespective of testing time point and children's grade level, shown in **Figure 4**.

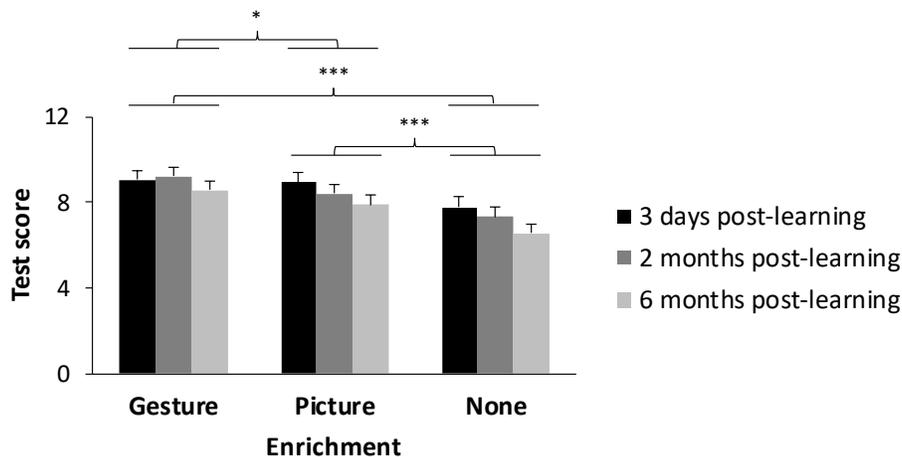

**Figure 4.** Test scores by learning condition and time point. Children demonstrated higher overall test scores following gesture-enriched learning and picture-enriched learning compared to non-enriched learning, and enrichment benefits did not significantly differ across time points. *$p$ < 0.05, ***$p$ < .001.

In agreement with previous studies in elementary school children and adults (Andrä et al., 2020; Macedonia & Knösche, 2011; Mayer et al., 2017), picture and gesture enrichment benefitted high school children's learning of both concrete and abstract word types compared to non-enriched learning: The mixed effects model indicated no significant interaction between learning condition and vocabulary type variables ($\chi2$ (2, $N$ = 75) = 1.08, $p$ = .58). Picture-enriched learning yielded significantly higher test scores than non-enriched learning for both concrete words ($\beta$ = 1.00, $t$ = 3.80, $p$ = .002) and abstract words ($\beta$ = 1.39, $t$ = 5.28, $p$ < .001), shown in **Figure 5**. Gesture-enriched learning also yielded significantly higher test scores than non-enriched learning for both concrete words ($\beta$ = 1.60, $t$ = 6.08, $p$ < .001) and abstract words ($\beta$ = 1.83, $t$ = 6.97, $p$ < .001).

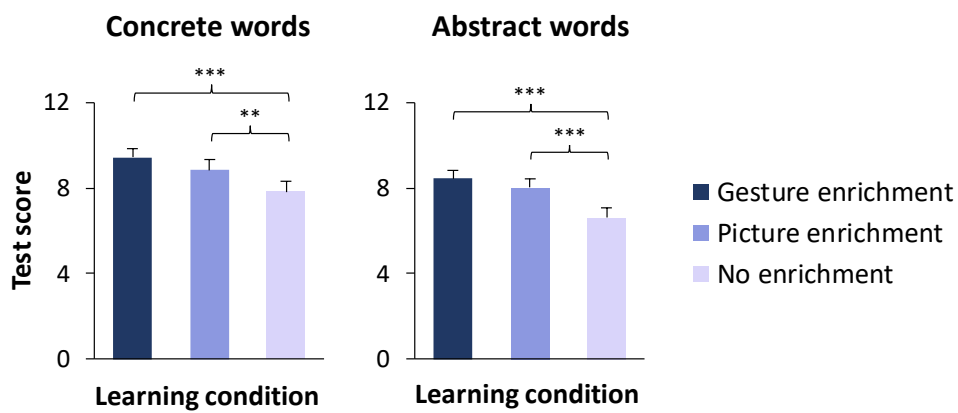

**Figure 5.** Test scores by learning condition and word type. Children demonstrated higher overall test scores following gesture-enriched learning and picture-enriched learning compared to non-enriched learning for both concrete words (left) and abstract words (right). **$p$ < 0.01, ***$p$ < .001.

The mixed modeling of children's test scores revealed several additional significant effects, which we report here for completeness. Test scores for concrete words were, overall, significantly higher than scores for abstract words, a main effect of word type ($\chi^2$ (1, $N$ = 75) = 44.45, $p$ < .001). There was also a significant main effect of time point ($\chi^2$ (2, $N$ = 75) = 25.57, $p$ < .001). These main effects were expected based on previous reports of children's greater performance for concrete than abstract nouns (Schwanenflugel, 1991), and reports of memory decay over time (Caramelli, Setti, & Maurizzi, 2004; Howe & Brainerd, 1989). The model also revealed significant grade × vocabulary and grade × time point interactions, shown in **Table 4**. Children's mean composite vocabulary test scores at 3 days, 2 months, and 6 months post-learning by condition are shown in **Table 5**.

|  |  | 3 days post-learning | | 2 months post-learning | | 6 months post-learning | |
|---|---|---|---|---|---|---|---|
|  |  | Grade 6 | Grade 8 | Grade 6 | Grade 8 | Grade 6 | Grade 8 |
|  |  | $M$ ($SE$) | $M$ ($SE$) | $M$ ($SE$) | $M$ ($SE$) | $M$ ($SE$) | $M$ ($SE$) |
| **Abstract words** | Picture | 8.95 (.73) | 8.00 (.52) | 9.18 (.59) | 7.22 (.58) | 8.38 (.61) | 6.14 (.46) |
|  | Gesture | 8.72 (.50) | 8.58 (.63) | 9.28 (.52) | 7.78 (.53) | 8.85 (.56) | 7.31 (.45) |
|  | None | 7.64 (.55) | 6.92 (.69) | 7.69 (.56) | 5.42 (.56) | 6.59 (.57) | 5.28 (.62) |
| **Concrete words** | Picture | 9.00 (.70) | 9.86 (.54) | 8.18 (.60) | 9.08 (.52) | 8.13 (.83) | 8.92 (.49) |
|  | Gesture | 7.90 (.60) | 10.97 (.57) | 9.38 (.61) | 10.39 (.63) | 8.23 (.61) | 9.89 (.60) |
|  | None | 7.00 (.70) | 9.53 (.60) | 7.77 (.64) | 8.50 (.61) | 6.33 (.69) | 8.06 (.55) |

**Table 5.** Children's composite scores on free recall and translation tests. $M$ = mean composite test score, $SE$ = standard error.

## Discussion

The present study was motivated by previous findings that adults'—but not elementary school children's—L2 vocabulary learning benefits to a greater extent from sensorimotor (gesture) than from multisensory (picture) enrichment (Andrä et al., 2020; Mathias et al., 2020; Mayer et al., 2015). We addressed the question of whether intermediate age groups would display enrichment benefits that are more comparable to those displayed by adults (i.e., gesture enrichment facilitating learning more than picture enrichment) or to those displayed by elementary school children (i.e., similar learning outcomes for gesture and picture enrichment). We found that both picture and gesture enrichment interventions were beneficial relative to non-enriched (auditory-only) learning for twelve-year-olds (sixth graders) and fourteen-year-olds (eighth graders). Interestingly, however, gesture-enriched learning was even more beneficial than picture-enriched learning for the eighth graders, while the sixth graders benefitted equivalently from learning enriched with pictures and gestures. This finding suggests that the effectiveness of gesture and picture enrichment techniques differs between younger and older L2 learners. While the pattern of enrichment effects for eighth graders qualitatively resembles that observed previously for young adults (Mathias et al., 2020; Mayer et al., 2015), the pattern of effects observed for sixth graders resembles that observed previously for elementary school children (Andrä et al., 2020). As was the case in previous studies on L2 enrichment, picture and gesture enrichment benefitted the learning of both concrete nouns (e.g., *tent*) and abstract nouns (e.g., *patience*), and effects of enrichment persisted over a long time scale (up to 6 months post-learning). Taken together, the findings suggest that congruent information presented in visual and motor modalities during auditory word learning may be differentially weighted by learners of different ages.

**Gesture Enrichment Benefitted Learning More than Picture Enrichment in Fourteen-Year-Old Children but not Twelve-Year-Old Children**

Children of both age groups were able to make use of enrichment information in a way that supported vocabulary knowledge. Across all time points and word types, performing gestures during L2 learning enhanced subsequent learning outcomes relative to auditory-only learning by about 22% in sixth graders and 25% in eighth graders. Viewing pictures enhanced learning outcomes by about 20% in sixth graders and 12% in eighth graders. These benefits are substantial when considering that children received minimal L2 exposure: Each L2 word was presented a total of only five times across three learning days. The children also never viewed the written words, and thus relied only on spoken words to form representations of the L2 tokens. Effects of enrichment were robust enough to support L2 translation for up to 6 months following learning. Beneficial effects of gesture enrichment on L2 vocabulary learning are consistent with a variety of psychological accounts, including embodied or grounded cognition perspectives (reviewed in Atkinson, 2010; Barsalou, 2008; Meteyard et al., 2012; Wellsby & Pexman, 2016), dual coding (Engelkamp & Zimmer, 1984; Hommel, Müsseler, Aschersleben, & Prinz, 2001; Paivio, 1991; Paivio & Csapo, 1969), imagery mechanisms (Jeannerod, 1995; Kosslyn, Thompson, & Ganis, 2006; Saltz & Dixon, 1982), and predictive coding accounts (Mayer et al., 2017; Mathias et al., 2020; von Kriegstein 2012).

The roughly equivalent benefits of gesture and picture enrichment for the sixth-grade children is consistent with the pattern of gesture and picture benefits recently shown in eight-year-old school children (Andrä et al., 2020). The superior effects of gesture enrichment relative to picture enrichment for the eighth-grade children is consistent with the pattern of gesture and picture benefits recently shown in adults (Mathias et al., 2020; Mayer et al., 2015; Repetto et al., 2017). Differences in enrichment benefits between age groups cannot be attributed to differences in gesture or picture stimuli, L2 perceptual characteristics, or training procedures, as these did not differ across age groups. Differences can also not be attributed to testing environments or to the translation from lab-based experiments in adults to a school setting, as both age groups in the current study were tested in similar school environments.

We offer two speculative explanations for the differences in effects of gesture and picture enrichment between sixth- and eighth-graders. The first explanation relates to potential advances in literacy in eighth graders compared to sixth graders. Children in the initial stages of reading skill acquisition may rely to a greater extent on visual context for L1 word learning relative to older children and adults (Nicholas & Lightbown, 2008). During the emergence of literacy, pictures and picture books serve as critical tools for language comprehension and vocabulary acquisition as they illustrate the meaning of spoken text (Ann Evans & Saint-Aubin, 2005; Feathers & Arya, 2012). Children are generally able to understand the referential nature of pictures—the idea that pictured contents can represent objects and concepts in the real world—by the age of two (Ganea, Allen, Butler, Carey, & DeLoache, 2009; Allen Preissler & Carey, 2004). While chapter books tend to include illustrations for children up to about 12 years, books intended for older children and adolescents rarely do so, and picture books tend not to be used as learning materials in older children's classrooms (Beckett, 2013). Instead, the majority of L1 vocabulary learning in adolescents and adults is thought to occur incidentally during the reading of written text (Webb, 2008); this is potentially also the case for L2 (Brown, Waring, & Donkaewbua, 2008; Grabe, 2009; Huckin & Coady, 1999). Thus, pictures are likely to play a greater role in aiding the learning of L2 vocabulary in younger children who may still be in the process of acquiring L1 competencies compared to older children (Spichtig et al., 2017).

The second explanation relates to differences in the degree to which children of different ages may rely on procedural and declarative memory systems for remembering L2 words. Theories of memory distinguish between procedural (implicit) and declarative (explicit) memory systems (Cohen, Poldrack, & Eichenbaum, 1997; Squire & Dede, 2015; Tulving & Madigan, 1970). Vocabulary learning is typically situated theoretically in the domain of declarative memory (Cabeza & Moscovitch, 2013), whereas other types of language learning such as grammar learning have become associated with the procedural memory system (Hamrick, 2015; Ullman, 2004). It has been suggested that gesture enrichment may engage the procedural memory system to a greater extent than audiovisual learning in adults (Macedonia & Mueller, 2016; Mathias et al., 2020), consistent with proposals that declarative and procedural memory systems in adults are interactive rather than distinct (Davis & Gaskell, 2009). Though declarative memory functions are not yet fully developed in younger children (Schneider, 2008), several studies have observed no differences between young children and adults in terms of procedural memory abilities (Finn et al., 2016; Karatekin, Marcus, & White, 2007; Meulemans, Ver der Linden, and Perruchet, 1998). It could be the case that, while children of both age groups made use of procedural memory systems for the learning of gesture-enriched vocabulary, picture enrichment recruited procedural memory systems only in the younger children. This would result in equivalent gesture- and picture-enriched learning outcomes in the younger children, and a reduction in picture enrichment benefits in the older children.

These potential explanations are currently speculative. Future studies may investigate how benefits of picture and gesture enrichment in children of different ages relate to the concurrent acquisition of reading and other academic skills, as well as the maturation of procedural and declarative memory. In terms of enrichment strategies that would be recommended for evidence-based L2 teaching, two open questions remain. First, does the current set of results extend to more commonly used forms of L2 instruction in which vocabulary acquisition is integrated into other L2 learning activities that are not focused explicitly on acquiring new vocabulary? Second, would the combination of gestures and pictures provide even larger enrichment benefits or would it create a dual attentional load resulting in inferior memory outcomes?

**Potential Brain Mechanisms Underlying Enhanced Learning Outcomes**

At present, the majority of neuroscience studies investigating learning enrichment have been conducted in adults. These studies suggest that beneficial effects of sensorimotor and multisensory enrichment derive, at least in part, from L2 representations stored in sensory and motor areas of the cortex. For example, listening to gesture-enriched L2 vocabulary elicits responses within regions associated with viewing and performing movements (Macedonia et al., 2011; Mayer et al., 2015), and these areas were found using a non-invasive neurostimulation method to causally facilitate the translation of L2 vocabulary (Mathias et al., 2020a, 2020b). These findings are comparable to neuroimaging studies in children, which have demonstrated preschoolers' greater motor (Kersey & James, 2013) and visual (James, 2010) cortical responses while viewing letters that they have previously been taught to write, compared to letters that they have been taught to recognize visually. Thus, the reactivation of neural sensory and motor structures at test that are involved in processing enrichment material during learning may drive enrichment benefits (multisensory learning theory, von Kriegstein, 2012; von Kriegstein & Giraud, 2006; Shams & Seitz, 2008).

**Conclusion**

We identified a dissociation in the effects of multisensory (picture) and sensorimotor (gesture) enrichment on L2 learning across twelve- and fourteen-year-old school children. Whereas fourteen-year-old children benefitted more from learning with gestures than with pictures, twelve-year-old children showed equivalent learning benefits following gesture- and picture-enriched learning. Gesture and picture enrichment strategies were tested systematically using large sample sizes of children in naturalistic school environments. We conclude that visual and motor enrichment information may be weighted differently by children of different ages, and that sensorimotor forms of enrichment may be more beneficial to older children for L2 vocabulary learning than audiovisual enrichment. The current findings provide evidence-based grounds for opting to include gestures rather than pictures in L2 vocabulary teaching for school children starting at fourteen years of age.

VOCABULARY LEARNING                                                                33Kuznetsova, A., Brockhoff, P. B., & Christensen, R. H. B. (2017). lmerTest package: tests in linear mixed effects models. *Journal of Statistical Software, 82*(13).

Lenth, R., Singmann, H., Love, J., Buerkner, P., & Herve, M. (2019). Package "emmeans": Estimated Marginal Means, aka Least-Squares Means. *The Comprehensive R Archive Network*, 1-67.

Li, J. T., & Tong, F. (2019). Multimedia-assisted self-learning materials: the benefits of E-flashcards for vocabulary learning in Chinese as a foreign language. *Reading and Writing, 32*(5), 1175-1195.

Luna, B., Garver, K. E., Urban, T. A., Lazar, N. A., & Sweeney, J. A. (2004). Maturation of cognitive processes from late childhood to adulthood. *Child Development, 75*(5), 1357-1372.

Macedonia, M. (2014). Bringing back the body into the mind: gestures enhance word learning in foreign language. *Frontiers in Psychology*, 5.

Macedonia, M., Bergmann, K., & Roithmayr, F. (2014). Imitation of a pedagogical agent's gestures enhances memory for words in second language. *Science Journal of Education, 2*(5), 162-169.

Macedonia, M., & Knösche, T. R. (2011). Body in mind: How gestures empower foreign language learning. *Mind, Brain, and Education, 5*(4), 196-211.

Macedonia, M., & Mueller, K. (2016). Exploring the neural representation of novel words learned through enactment in a word recognition task. *Frontiers in Psychology, 7*, 953.

Mahmoudi, S., Jafari, E., Nasrabadi, H. A., & Liaghatdar, M. J. (2012). Holistic Education: An Approach for 21 Century. *International Education Studies, 5*(2), 178-186.

*Mathias, B., *Sureth, L., Hartwigsen, G., Macedonia, M., Mayer, K. M., & von Kriegstein, K. (2020a). Visual sensory cortices causally contribute to auditory word recognition following sensorimotor-enriched vocabulary training. *Cerebral Cortex, 31*(1), 513-528. *Joint first authors

VOCABULARY LEARNING 35Oxford, R., and Crookall, D. (1990). Vocabulary learning: A critical analysis of techniques. *TESL Canada*, 7, 9-30.

Paivio, A. (1991). Dual coding theory: Retrospect and current status. *Canadian Journal of Psychology, 45*, 255–287.

Paivio, A., & Csapo, K. (1969). Concrete image and verbal memory codes. *Journal of Experimental Psychology, 80*(2p1), 279.

Prince, M. (2004). Does active learning work? A review of the research. *Journal of Engineering Education, 93*(3), 223-231.

Raviv, L., & Arnon, I. (2018). The developmental trajectory of children's auditory and visual statistical learning abilities: modality-based differences in the effect of age. *Developmental Science, 21*(4), e12593.

Repetto, C., Pedroli, E., & Macedonia, M. (2017). Enrichment effects of gestures and pictures on abstract words in a second language. *Frontiers in Psychology, 8*, 2136.

Sadoski, M., & Paivio, A. (2013). *Imagery and text: A dual coding theory of reading and writing*. New York: Routledge.

Saffran, J. R., Johnson, E. K., Aslin, R. N., & Newport, E. L. (1999). Statistical learning of tone sequences by human infants and adults. *Cognition, 70*(1), 27-52.

Saltz, E., & Dixon, D. (1982). Let's pretend: The role of motoric imagery in memory for sentences and words. *Journal of Experimental Child Psychology, 34*(1), 77-92.

Sambanis, M. (2013). *Fremdsprachenunterricht und Neurowissenschaften*. Narr Francke Attempto Verlag.

Schneider, W. (2008). The development of metacognitive knowledge in children and adolescents: Major trends and implications for education. *Mind, Brain, and Education, 2*(3), 114-121.

Schmitt, N., & Schmitt, D. (2020). *Vocabulary in language teaching*. Cambridge, UK: Cambridge University Press.

VOCABULARY LEARNING                                                                 36Schwanenflugel, P. J. (1991). Why are abstract concepts hard to understand? In P.J. Schwanenflugel (Ed.), *The psychology of word meanings* (pp. 223-250). Hillsdale, N.J.: LEA.

Silverman, R., & Hines, S. (2009). The effects of multimedia-enhanced instruction on the vocabulary of English-language learners and non-English-language learners in pre-kindergarten through second grade. *Journal of Educational Psychology, 101*(2), 305.

Shams, L., & Seitz, A. R. (2008). Benefits of multisensory learning. *Trends in Cognitive Sciences, 12*(11), 411-417.

Spichtig, A., Pascoe, J., Ferrara, J., & Vorstius, C. (2017). A comparison of eye movement measures across reading efficiency quartile groups in elementary, middle, and high school students in the US. *Journal of Eye Movement Research, 10*(4), 5.

Tulving, E., & Madigan, S. A. (1970). Memory and verbal learning. *Annual Review of Psychology, 21*(1), 437-484.

Squire, L. R., & Dede, A. J. (2015). Conscious and unconscious memory systems. Cold Spring Harbor Perspectives in Biology, 7(3), a021667.

von Kriegstein, K., & Giraud, A. L. (2006). Implicit multisensory associations influence voice recognition. *PLoS biology, 4*(10), e326.

von Kriegstein, K. (2012). A multisensory perspective on human auditory communication. In M. M. Murray, M. T. Wallace (Eds.), *The neural bases of multisensory processes* (pp. 683-700). Boca Raton: CRC Press/Taylor & Francis.

Webb, S. (2008). The effects of context on incidental vocabulary learning. *Reading in a Foreign Language, 20*(2), 232-245.

Wellsby, M., & Pexman, P. M. (2014). Developing embodied cognition: Insights from children's concepts and language processing. *Frontiers in Psychology, 5*, 506.

Zemlock, D., Vinci-Booher, S., & James, K. H. (2018). Visual–motor symbol production facilitates letter recognition in young children. *Reading and Writing, 31*(6), 1255-1271.

**Supplementary Material**

**Table S1.** Linear mixed effects regression coefficient estimates for fixed effects of enrichment type, grade level, time point, and word type. *SD* = standard deviation, *SE* = standard error, CI = 95% confidence interval.

| *Random effects* | | |
|---|---|---|
| | *Variance* | SD |
| Participant Intercept | 5.44 | 2.33 |
| Vocabulary Slope | .05 | .22 |

| *Fixed effects* | | | | |
|---|---|---|---|---|
| | Estimate | *SE* | *t* | CI |
| Intercept | 8.19 | .29 | 28.08 | 7.62, 8.77 |
| Learning [Picture] | −.52 | .18 | −2.85 | −.88, −.16 |
| Learning [No Enrichment] | −1.71 | .18 | −9.35 | −2.07, −1.35 |
| Grade | .03 | .58 | .06 | −1.13, 1.19 |
| Vocabulary | 1.07 | .15 | 7.04 | .77, 1.37 |
| Time point [2 months] | −.91 | .18 | −4.98 | −1.27, −.55 |
| Time point [6 months] | −.26 | .18 | −1.44 | −.62, .10 |
| Learning [Picture] × Grade | −.86 | .37 | −2.36 | −1.58, −.14 |
| Learning [No Enrichment] × Grade | −.31 | .37 | −.86 | −1.03, .40 |
| Learning [Picture] × Vocabulary | −.15 | .37 | −.42 | −.87, .56 |
| Learning [No Enrichment] × Vocabulary | .23 | .37 | .64 | −.48, .95 |
| Grade × Vocabulary | 2.88 | .30 | 9.50 | 2.29, 3.48 |
| Learning [Picture] × Time [2 months] | −.58 | .45 | −1.29 | −1.46, .30 |
| Learning [No Enrichment × Time [2 months] | −.73 | .45 | −1.63 | −1.61, .15 |
| Learning [Picture] × Time [6 months] | −.69 | .45 | −1.55 | −1.57, .18 |
| Learning [No Enrichment] × Time [6 months] | −.59 | .45 | −1.32 | −1.47, .29 |
| Grade × Time [2 months] | −.92 | .37 | −2.53 | −1.64, −.21 |
| Grade × Time [6 months] | −1.29 | .37 | −3.52 | −2.01, −.57 |

VOCABULARY LEARNING                                                                 39| | | | | |
|---|---|---|---|---|
| Vocabulary × Time [2 months] | .26 | .37 | .70 | −.46, .97 |
| Vocabulary × Time [6 months] | .21 | .37 | .57 | −.51, .93 |
| Learning [Picture] × Grade × Vocabulary | −.39 | .73 | −.54 | −1.83, 1.56 |
| Learning [No Enrichment] × Grade × Vocabulary | .12 | .73 | .17 | −1.31, 1.56 |
| Learning [Picture] × Grade × Time [2 months] | .74 | .90 | .83 | −1.02, 2.50 |
| Learning [No Enrichment] × Grade × Time [2 months] | .71 | .90 | .80 | −1.05, 2.47 |
| Learning [Picture] × Grade × Time [6 months] | 1.25 | .90 | 1.39 | −.51, 3.01 |
| Learning [No Enrichment] × Grade × Time [6 months] | .05 | .90 | .05 | −1.71, 1.81 |
| Learning [Picture] × Vocabulary × Time [2 months] | .09 | .90 | .10 | −1.67, 1.85 |
| Learning [No Enrichment] × Vocabulary × Time [2 months] | .08 | .90 | .09 | −1.68, 1.84 |
| Learning [Picture] × Vocabulary × Time [6 months] | −1.11 | .90 | −1.24 | −2.87, .65 |
| Learning [No Enrichment] × Vocabulary × Time [6 months] | .02 | .90 | .03 | −1.74, 1.78 |
| Grade × Vocabulary × Time [2 months] | .32 | .73 | .44 | −1.11, 1.76 |
| Grade × Vocabulary × Time [6 months] | .03 | .73 | .04 | −1.41, 1.46 |
| Learning [Picture] × Grade × Vocabulary × Time [2 months] | 1.21 | .79 | .67 | −2.31, 4.73 |
| Learning [No Enrichment] × Grade × Vocabulary × Time [2 months] | −.21 | .79 | −.12 | −3.73, 3.12 |
| Learning [Picture] × Grade × Vocabulary × Time [6 months] | 1.72 | .79 | .96 | −1.80, 5.24 |
| Learning [No Enrichment] × Grade × Vocabulary × Time [6 months] | .46 | .79 | .25 | −3.07, 3.97 |